\documentclass[a4paper, 12pt]{article}
\usepackage{cite}
\usepackage{rotating}
\usepackage[english]{babel}
\usepackage[a4paper, inner=2cm, outer=2cm, top=3cm, bottom=3cm]{geometry}
\usepackage[dvipsnames]{xcolor}
\usepackage{mathtools}
\usepackage{pstricks}
\usepackage{caption}
\usepackage{graphicx}
\usepackage{amsmath}
\usepackage{amssymb}
\usepackage{mathcomp}
\usepackage{textcomp}
\usepackage{enumitem}
\usepackage{physics}
\usepackage{romannum}
\usepackage[doublespacing]{setspace}
\usepackage{subcaption}
\usepackage{hyperref}

\hypersetup{
  colorlinks   = true,
  urlcolor     = Blue, 
  linkcolor    = Black,
  citecolor    = Black
}

\title{On the Quantum Improved Affine Gravity}
\author{Raihaneh Moti and Ali Shojai\\
\textit{\small Department of Physics, University of Tehran, Tehran, Iran}}
\date{}

\newcommand{\ArcTan}{\text{Tan}^{-1}}
\newcommand{\thickhat}[1]{\mathbf{\hat{\text{$#1$}}}}

\begin{document}
\maketitle
\pagenumbering{arabic}
\begin{abstract}
Improvement of the classical gravity with the running gravitational coupling obtained from asymptotically safe gravity, is a good way of considering the effects of quantum gravity. This is usually done for metric theories of gravity. Here we investigate the effects of such an improvement for pure affine theories of gravity. To motivate the approach, we first consider the effects of quantum improvement on the connection using metric theory and investigate the effects on the causal structure of black hole solution. Next in the framework of Schr\"odinger--Eddington affine theory, the general way of affine improvement is presented and a spherically symmetric solution is obtained and compared with other ways of improvement.
\end{abstract}

\section{Introduction}
It is known that, besides the Einstein theory of General Relativity, which considers metric as a dynamical object to describe the gravity, there are other theories based on the assumption of independence of the affine connection and metric, considering it as an independent dynamical object\cite{Schrodinger-B,Eddington-B,E-S,Kijowski}. 
Although the modern methods based on the gauge theory are a more proper way for the unification of various interactions, it is undeniable that the affine gravity was one of the primary efforts for an explanation of the electromagnetism in the covariant formalism suggested by Einstein. Affine theory is followed by considering more symmetries in the recent years\cite{EM-GR1,EM-GR2,EM-GR4,EM-GR3}, however none of them achieves a considerable status among the modern unification methods yet. After all, the main physical point about affine theories is the use of connection instead of metric as the dynamical variable.

The affine formalism becomes notable when we encounter quantum gravity models relying on the quantum field theory implement, because of the fundamental nature of the affine connection. The main feature of these theories is the usage of a gauge field to describe gravity. 
For gravitational interaction, what describes the gravity field is in fact the measure of rotation of local inertial frame at each point with respect to the neighboring ones, i.e. the deviation from Euclidean parallel transport. Clearly, this deviation is described by the affine connection ($\Gamma^{\gamma}_{\mu\nu}$) in the curved space--times, and the affine connection is precisely the gravity gauge field. This is a good reason to attend more to non--metric theories in some quantum gravity models.

In general, depending on the chosen fundamental dynamical variables there are three classes of gravity theory. The \textit{pure metric formulation}, based on the consideration of the metric as the independent variable\cite{PM,PM2}. This choice with the simplest possible action, the Einstein--Hilbert action, results in the equivalence of the affine connection and the Levi--Civita connection $\genfrac{\lbrace}{\rbrace}{0pt}{}{\gamma}{\mu\nu}$. 

Choosing the connection in addition to the metric as an independent dynamical variable, leads to the second class which is known as \textit{metric--affine formulation}\cite{MA1,MA2,MA3,MA4}. And, at last, \textit{pure affine theories} are the ones for which the affine connection is the only fundamental dynamical variable\cite{Schrodinger-B,Eddington-B,E-S,Kijowski,EM-GR1,EM-GR2,EM-GR3,EM-GR4,PA1,PA2} (See references in \cite{EM-GR1} for detailed reviews). 

Since the affine is the gauge field for gravity, the pure affine gravity may be ideally suited for the quantization process. One way to examine this idea is to study the quantum improvement of such theories in the context of the \textit{asymptotic safety conjecture}.

According to the asymptotic safety conjecture, any quantum field theory which ends to a non--Gaussican UV fixed point, lying in the finite dimensional tangent space, is safed from UV divergences, and becomes predictable\cite{Weinberg1,Weinberg2,Becker,Niedermaier}. There are various evidences that the gravity flow has a non--Gaussian fixed point and is asymptotically safe (See references in \cite{NGF}). Using the functional renormalization group methods to find this non--Gaussian fixed point, the behavior of the gravitational running coupling could be derived\cite{Reuter-1st}.

Although it is expected that studying the effects of this quantization method needs the solution of the exact renormalization group equation, i.e. effective average action, but it seems too complicated to derive it formally if not impossible. Thus, it is claimed that the improvement of coupling constant $g_0$ in the classical theory to the running one $g_k$, can \textit{hopefully} restore the quantum effects at least to some extent\cite{Reuter & Weyer,Reuter & Saueressig}.

The improvement can be done through different strategies as it is discussed in \cite{Reuter & Weyer,our3}. There, it is shown, that the most physical way for quantum improvement of the field equations, is the action improvement using the cutoff identification $k=\xi_0/\chi$ as a scale parameter. $\xi_0$ is a dimensionless constant and $\chi$ is a length dimensional quantity. In a curved space--time, the equivalence principle and geodesic deviation concept gives a natural lenght scale, which is given by the curvature tensor and describes the neighbourhood of a point in which the freely falling frame (at the point) is inertial. As it is discussed in\cite{our3} a proper function of curvature invariants ($\chi(\chi_i)$, where $\chi_i$'s are the curvature invariants) is the best choice for our purpose. Unfortunately one can not fix a unique form for this function using this proposal, but various assumptions such as energy conditions\cite{our4}, singularities and etc. can be used to restrict the possible forms.

The effects of various strategies of the improvement of the gravitational coupling constant on the running one and the cutoff identification are studied for a wide range of problems like cosmological solutions \cite{NGF,cos1,cos2,cos3,our1}, black holes \cite{Becker,Bonanno & Reuter,BH1,BH2,BH3,our2}, energy conditions\cite{our4,our5}, vacuum fluctuations\cite{vac1,vac2,vac3} and so on. 

The functional renormalization group methods besides other assumptions \cite{Reuter-1st, Souma}, leads to the antiscreening running gravitational coupling as
\begin{equation} \label{RGC}
  G(\chi) = \frac{G_0}{1+f(\chi)}
\end{equation}
where $f(\chi) \equiv \xi /\chi^2$.
The small scaling constant $\xi$ equals to $G_0 \omega_q \xi_0 $  provided that the reference constant $ G_0 $ be the experimentally observed value of Newton's constant $G_N$ and $\omega_q = \frac{4}{\pi}(1-\frac{\pi^{2}}{144})$ \cite{Bonanno & Reuter}.

The above mentioned improvements are done in metric theories. But what about affine theories? Since the affine connection is in fact the \textit{gravity force}, it is more reasonable to improve it in the classical theory by quantum effects.

Here we are looking for the effects of such an improvement. For this, we would consider the affine connection as the main dynamical object of the theory, using the affine gravity theories.

First in section II, we present a simple way for introducing the quantum improvements of asymptotically safe gravity method in the affine connection in the framework of metric gravity, and investigate its effects on the causal structure of black hole solution. Then, in section III, the general structure of the improved affine gravity is investigated in the framework of Schr\"odinger--Eddington affine theory.

\section{Motivations from metric theory} \label{II}
Before looking for a complete model for quantum improved affine gravity, let us give some motivations and illustrations that can be obtained from the metric theory.
\subsection{The method of affine improvement}
Here we would formulate a way that through it one can obtain an improved connection. This needs attention to the Newtonian limit of the gravitational theory.
Consider the geodesic equation 
\begin{equation}
\dv[2]{x^{\gamma}}{s} + \genfrac{\lbrace}{\rbrace}{0pt}{}{\gamma}{\mu\nu} \dv{x^{\mu}}{s}\dv{x^{\nu}}{s} = 0 \ .
\end{equation}

The connection contains two type of terms. Terms related to the use of curvilinear coordinates and terms representing the gravitational force. Let us to forget the first terms for now (e.g. by going to a coordinate system in which the connection is only the result of gravity). In this way the term $\genfrac{\lbrace}{\rbrace}{0pt}{}{\gamma}{\mu\nu} \dv{x^{\mu}}{s}\dv{x^{\nu}}{s}$ is proportional to the gravity field. As a result it is proportional to $G_0$, as it should have Newtonian limit\footnote{It has to be noted that for purely gravitational solutions, like black holes or free gravitational waves, this conclusion may breaks. But in general one can always go to the Newtonian limit and relate the parameters in the solution to Newton's constant.}.

On the other hand, the theory of general relativity is a metric compatible and torsion--free theory. Then, for this theory, $\genfrac{\lbrace}{\rbrace}{0pt}{}{\gamma}{\mu\nu} = \Gamma^{\gamma}_{\mu\nu}$, and thus
\begin{equation}
\Gamma^{\gamma}_{\mu\nu}\dv{x^{\mu}}{s}\dv{x^{\nu}}{s} \propto G_0 \ .
\end{equation}
Now, we can \textit{propose} that the improved affine connection, $\thickhat{\Gamma}^{\gamma}_{\mu\nu}$ is given by
\begin{equation} \label{Im-Aff}
\thickhat{\Gamma}^{\gamma}_{\mu\nu}(\mathcal{G}) = \frac{G(\chi)}{G_0} \genfrac{\lbrace}{\rbrace}{0pt}{}{\gamma}{\mu\nu}_{g^{NI}} \ .
\end{equation}
As a result of the antiscreening behavior \eqref{RGC}, $G(\chi)/G_0 < 1$, and the quantum improvement weakens the gravitation attraction. This would show itself as  compactification of light cones, and would be illustrated in the next subsection.

It has to be noted that this in general, describes a space--time with metric $\mathcal{G}_{\alpha\beta}$ (which should be determined) with non--metricity and torsion (See references \cite{Ad-1,Ad-2,Ad-3} for more on the gravity with torsion. \textit{That is to say, quantum improvement of the connection can in general introduce torsion and/or non--metricity to the non--improved space--time (i.e. the classical space--time without any quantum correction).} This means that one may observes the effects of quantum corrections of a Riemann space--time within the geometrical elements of a general classical affine manifold such as torsion and/or non-metricity. If such a proposal is going to be a good one, \textit{we have to} start from a classical theory with torsion and non-metricity and see how the quantum effects alters them. This is what we shall do in the next section, but for now let us to go further to get a taste of how this can happen.

For this, we should solve \eqref{Im-Aff} for the metric $\mathcal{G}_{\alpha\beta}$ using the symmetries of the non--improved metric $g^{NI}_{\alpha\beta}$.
As it is stated earlier, there is a notable point in the practical process of the affine improvement. Not all of the $\genfrac{\lbrace}{\rbrace}{0pt}{}{\gamma}{\mu\nu}$'s are the result of the gravitational interaction, rather, some of them are the signs of using a curvilinear coordinate system. 
It is clear that setting the source of gravitational field zero, what remains in the connection is the effects of using curvilinear coordinates.

Thus, one way to study the affine improvement, is just to consider only the gravitational parts, i.e. those ones which are resulted from the rotations of the local basis by gravity, $\genfrac{\lbrace}{\rbrace}{0pt}{}{\gamma}{\mu\nu}^{Grav.}_{g^{NI}}$.
After improving these portions, we may add them to their curvilinear counterparts, $\genfrac{\lbrace}{\rbrace}{0pt}{}{\gamma}{\mu\nu}^{Coor.}_{g^{NI}}$, to obtain $\thickhat{\Gamma}^{\gamma}_{\mu\nu}(\mathcal{G})$. In other worlds,
\begin{equation}
\thickhat{\Gamma}^{\gamma}_{\mu\nu}(\mathcal{G}) =  \frac{G(\chi)}{G_0} \genfrac{\lbrace}{\rbrace}{0pt}{}{\gamma}{\mu\nu}^{Grav.}_{g^{NI}}+ \genfrac{\lbrace}{\rbrace}{0pt}{}{\gamma}{\mu\nu}^{Coor.}_{g^{NI}} \ .
\label{main}
\end{equation}

One may argue that in general relativity it is not possible to separate gravity from coordinate system curvature. But it should be noted that what is meant in the above statement is simply that for solutions for which it is possible to remove the gravity effects (and get
a flat space–time) by setting some parameters to zero (like the Schwarzschild solution) what remains in the connection is the results of using curvilinear coordinates for flat space–time.

In the next subsection we shall illustrate such a process by an example. As it is stated before one should start from a classical theory with torsion and non-metricity, but in what follows in this section we shall only present some simple calculations to get motivated that the quantum effects can be reorganized in terms of torsion and non-metricity. Then in section III, we shall start from a general space--time and present the full model.

\subsection{An example: Spherically symmetric vacuum solution}
As an example, let's improve a static spherically symmetric space--time, using the proposed method. Such a space--time is described by
\begin{equation} \label{SSM0}
g_{tt}=A(r),\quad\quad g_{rr}=-B(r),\quad\quad g_{\theta\theta}=-r^2,\quad\quad g_{\varphi\varphi}=-r^2 \sin^2\theta \ . 
\end{equation}
Then the non--zero components of the connection are
\begin{align}
& \genfrac{\lbrace}{\rbrace}{0pt}{}{r}{tt}(g) =\frac{A'}{2B} && \genfrac{\lbrace}{\rbrace}{0pt}{}{t}{rt}(g) =\frac{A'}{2A} &&\genfrac{\lbrace}{\rbrace}{0pt}{}{r}{rr}(g) =\frac{B'}{2B}  \nonumber \\
& \genfrac{\lbrace}{\rbrace}{0pt}{}{\theta}{r\theta}(g) =\frac{1}{r} && \genfrac{\lbrace}{\rbrace}{0pt}{}{r}{\theta\theta}(g) =\frac{-r}{B} &&\genfrac{\lbrace}{\rbrace}{0pt}{}{\theta}{\theta\varphi}(g) =\cot\theta  \nonumber \\
& \genfrac{\lbrace}{\rbrace}{0pt}{}{\varphi}{r\varphi}(g) =\frac{1}{r} &&\genfrac{\lbrace}{\rbrace}{0pt}{}{r}{\varphi\varphi}(g) =\frac{-r}{B}\sin^2\theta &&\genfrac{\lbrace}{\rbrace}{0pt}{}{\theta}{\varphi\varphi}(g) =-\cos\theta\sin\theta
\end{align}
where we have used the notations $A(r)\equiv A$ and $B(r)\equiv B$, a prime over any quantity to denote the derivative with respect to its argument, here $r$.

The portions related to the use of curvilinear coordinate,  are derived by setting $A=B=1$. On using this and equations (\ref{RGC}) and (\ref{main}), the quantum improved connection has the following non--vanishing components for our spherical static space--time
\[
\thickhat{\Gamma}^{t}_{tr} = \frac{1}{1+f} \frac{A'}{2A} \quad,\quad \thickhat{\Gamma}^{r}_{rr} = \frac{1}{1+f} \frac{B'}{2B} \]
\begin{equation}
\thickhat{\Gamma}^{r}_{tt} = \frac{1}{1+f} \frac{A'}{2B} \quad,\quad \thickhat{\Gamma}^{r}_{\theta\theta} =-r+\frac{r}{1+f} (1-\frac{1}{B}) \quad,\quad \thickhat{\Gamma}^{r}_{\varphi\varphi} =(-r+\frac{r}{1+f} (1-\frac{1}{B}))\sin^2\theta \label{IMC-1}
\end{equation}

On the other hand, assuming that the improvement does not change the spherical symmetry of the space--time, we can set 
\begin{equation}
\mathcal{G}_{tt}=X(r),\quad\quad \mathcal{G}_{rr}=-Y(r),\quad\quad \mathcal{G}_{\theta\theta}=-r^2,\quad\quad \mathcal{G}_{\varphi\varphi}=-r^2 \sin^2\theta
\end{equation}
for the improved metric and 
\[
\thickhat{\Gamma}^{t}_{tr} =  \frac{X'}{2X} +K^{t}_{tr} + L^{t}_{tr} \quad,\quad \thickhat{\Gamma}^{r}_{rr} = \frac{Y'}{2Y}+K^{r}_{rr} + L^{r}_{rr} \]
\begin{equation}
\thickhat{\Gamma}^{r}_{tt} =  \frac{X'}{2Y}+K^{r}_{tt} + L^{r}_{tt} \quad,\quad \thickhat{\Gamma}^{r}_{\theta\theta} =\frac{-r}{Y}+K^{r}_{\theta\theta} +L^{r}_{\theta\theta} \quad,\quad \thickhat{\Gamma}^{r}_{\varphi\varphi} =\frac{-r}{Y} \sin^2\theta +K^{r}_{\varphi\varphi}+L^{r}_{\varphi\varphi} \label{IMC-2}
\end{equation}
for the improved components of the connection, 
where $K^{\gamma}_{\mu\nu}$ is the contorsion tensor and $L^{\gamma}_{\mu\nu}$ is the deformation caused by the non--metricity.

In order to go further for this simple example, let us adopt the naive cutoff identification $f=\xi/r^2$ for simplicity. Equations \eqref{IMC-1} and \eqref{IMC-2} are a set of 5 equations for 12 independent unknown components of the metric $\mathcal{G}_{\mu\nu}$, the contorsion tensor $K^{\gamma}_{\mu\nu}$  and the deformation $L^{\gamma}_{\mu\nu}$. To have a solution, we restrict ourselves to a subspace of possible solutions that have $K^{\gamma}_{\mu\nu}=0$. Indeed, for more convenience, we assume that the $ L^{t}_{tr}=L^{r}_{rr}=0$, although the results can be extended to more general solutions. These considerations leave us with two differential equations for $X(r)$ and $Y(r)$, and three algebraic equations for $L^{r}_{tt}$, $L^{r}_{\theta\theta}$ and $L^{r}_{\varphi\varphi}$.

Nothing that the non--improved solution is the Schwarzschild vacuum solution, for which $A(r)=-B(r)^{-1} = 1-r_s/r$, the improved metric and deformation components are given by
\begin{align}
& \mathcal{G}_{tt}= e^{(\zeta\ArcTan\frac{u}{\zeta}+\ln{\frac{(1-u)}{\sqrt{u^2+\zeta^2}}})/(1+\zeta^2)} \label{Im-tt}\\
& \mathcal{G}_{rr}= -e^{-(\zeta\ArcTan\frac{u}{\zeta}+\ln{\frac{(1-u)}{\sqrt{u^2+\zeta^2}}})/(1+\zeta^2)} \label{Im-rr}\\
& \mathcal{G}_{\theta\theta}=-u^2 \label{Im-thth}\\
& \mathcal{G}_{\varphi\varphi}=-u^2 \sin^2\theta \label{Im-phph}\\
& \tilde{L}^{r}_{tt} = \frac{(u^2+\zeta^2)^{-(2+\zeta^2)/(1+\zeta^2)}}{2\abs{-1+u}u} \left( e^{2\zeta\frac{\ArcTan\tfrac{u}{\zeta}}{1+\zeta^2}} \abs{1-u}^{\frac{2}{1+\zeta^2}} u^2 - (u-1)^2 (u^2+\zeta^2)^{\frac{1}{1+\zeta^2}} \right) \\
& \tilde{L}^{r}_{\theta\theta} = \tilde{L}^{r}_{\varphi\varphi}/\sin^2\theta = u - \frac{u^2}{u^2+\zeta^2} - u e^{(\zeta\ArcTan\frac{u}{\zeta}+\ln{\frac{(1-u)}{\sqrt{u^2+\zeta^2}}})/(1+\zeta^2)}
\end{align}
where $u \equiv r/r_s$, $\zeta \equiv \xi/r_s^2$ and $\tilde{L}^{\gamma}_{\mu\nu} \equiv L^{\gamma}_{\mu\nu}/r_s $ are dimensionless quantities.

The causal structure of such an space--time can be studied in two steps. First, we shall look for its light cones, and then using Carter--Penrose diagrams the causal structure is studied.

\subsubsection{Light cones}
The light cone is the local structure, that restricts the possible causal interactions by defining a tangent space at any point of the space--time. This causal structure is specified by the constant of causal structure $c_{ST}$, where in general relativity equals to the $c_0 =3 \times 10^8 \ ms^{-1}$. 
Although $c_0$ is known as the speed of light, as categorized
by Ellis and Uzan\cite{Ellis}, there are four different interpretations for $c_0$: The electromagnetic wave velocity ($c_{EM}$), the constant of space--time structure ($c_{ST}$), the gravitational wave velocity ($c_{GW}$) and the space--time--matter coupling ($c_E$). Since the custom viewpoint about the equivalence of the four dimensionless quantities $c_{EM}/c_0$, $c_{ST}/c_0$, $c_{GW}/c_0$ and $c_E/c_0$, is based on the standard theories of the gravity and electromagnetic with their classical limit, any modification of these theories, such as the quantum improvement, may lead to defeat of these equivalences\cite{Izadi}.

\begin{figure}
\rotatebox{90}{\includegraphics[width=0.9\textheight]{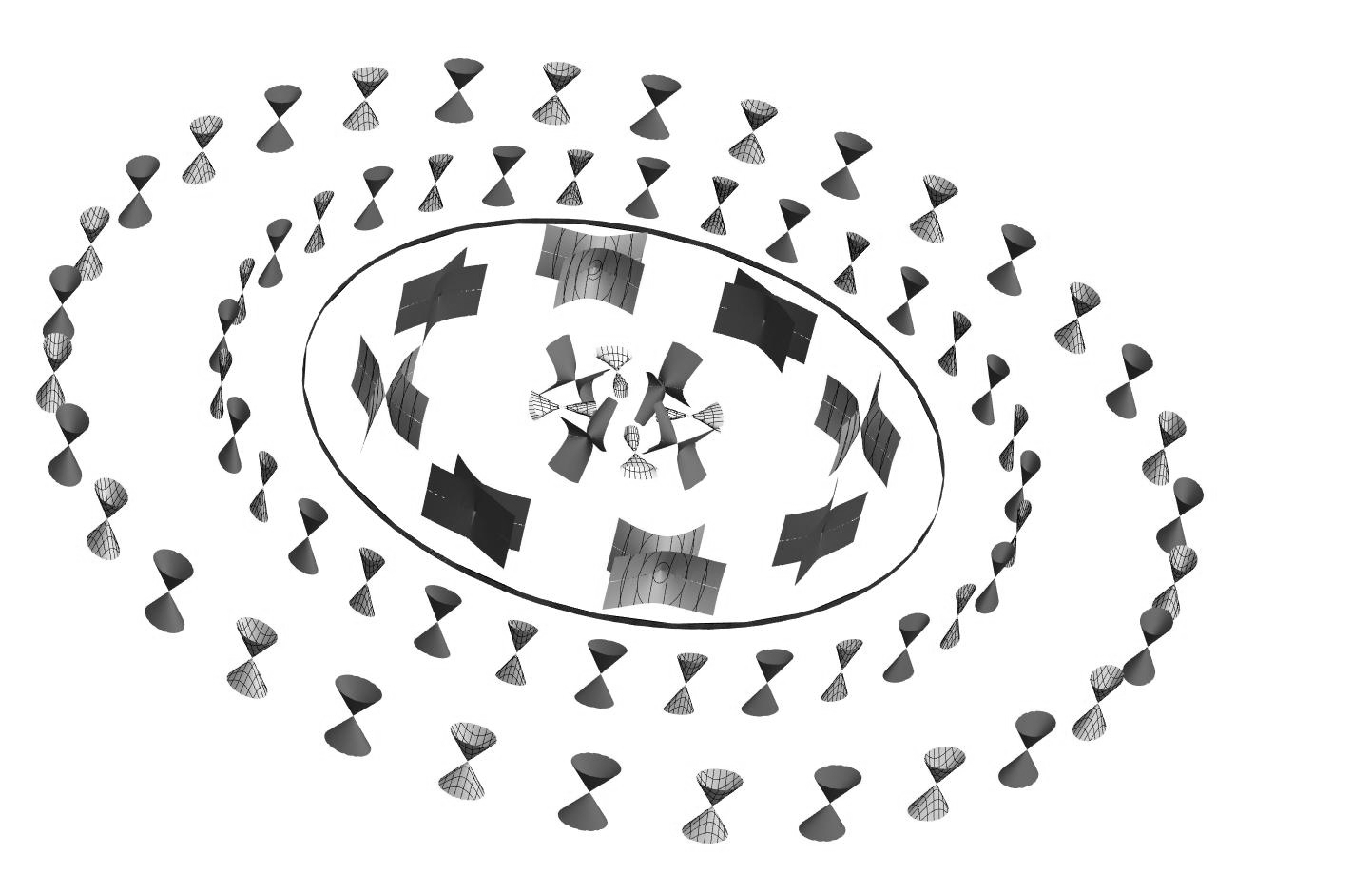}}
\caption{The meshed surfaces, denotes the improved light cones, while the grey ones, are the classical non--improved causal light cones. Clearly, the polar symmetries are saved, but by decreasing the radial distance, the cones are more compactified by the quantum effects.}
\label{NC}
\end{figure}

In order to find $c_{ST}$, we should expand the classical metric $g^{NI}$, in the local basis of the improved metric $\mathcal{G}$. Thus, we should find
\begin{equation}
\tilde{g}_{ij}=\tilde{e}^a_i(\mathcal{G}) \tilde{e}^b_j(\mathcal{G}) g_{ab}^{NI}
\end{equation}
where 
\begin{equation}
\tilde{e}^a_l(\mathcal{G}) = \left(\frac{1}{\sqrt{\mathcal{G}_{tt}}},\sqrt{\mathcal{G}_{tt}},\frac{1}{r},\frac{1}{r\sin\theta} \right) 
\end{equation}
are the tetrad basis of the improved space--time. Then, for a non--improved space--time \eqref{SSM0},  the line element of the $\tilde{g}_{ij}$ becomes
\begin{equation}
\dd{s}^2 = \frac{\mathcal{G}_{tt}}{1-r_s/r} \left( (\frac{1-r_s/r}{\mathcal{G}_{tt}})^2 \dd{t}^2- \dd{r}^2 - r^2 \dd{\theta}^2 - r^2 \sin\theta^2\dd{\varphi}^2  \right) \ .
\end{equation}
Now considering a radial null ray, we will find the constant of causal structure as
\begin{equation}
c_{ST} = \abs{\frac{1-r_s/r}{\mathcal{G}_{tt}}} c_0 \ .
\end{equation}
Since $\mathcal{G}_{tt} > 1-r_s/r $, then $c_{ST} <c_0$. And, this would result in the compactification of the light cone, as can be seen in figure (\ref{NC}). One can see that the improved light cones (meshed ones) would be compactified by going forward towards the classical singularity. This compactification is the sign of a decrease in the convergence behavior of the null geodesics, which is exactly what we anticipate from the antiscreening gravitational running coupling, equation \eqref{RGC}.

\subsubsection{Penrose--Carter diagram}
To study the causal structure of a black hole solution, it is better to take a look at its Penrose--Carter diagram, as it represents the entire space--time, by making the infinities and singularities accessible. The first step to draw Penrose--Carter diagram for our solution is making the coordinates to have a finite region\cite{Poisson}. Going to the plane $(\theta=\text{const.},\varphi=\text{const.})$, one can get the conformally flat metric
\begin{equation}
\dd{s}^2 = X(u) \left(\dd{t}^2-\dd{R(u)}^2 \right))
\end{equation}
where
\begin{equation}
R(u) = \int^u_1 \frac{\dd{u}}{X(u)}  \ .
\end{equation}
If we use the compactification
\begin{align}
\label{comp1}
& x = \tfrac{1}{2} \left[\tanh(R(u)+t) - \tanh(R(u)-t) \right] \\
\label{comp2}
& y = - \tfrac{1}{2} \left[\tanh(R(u)+t) + \tanh(R(u)-t) \right] \ ,
\end{align}
then by determining the boundaries of these coordinates, the diagram would be determined.

For our improved curved space--time $\mathcal{G}$, the $R(u)$ should be driven from the $X(u)=\mathcal{G}_{tt}$.
Since, the exact solution of $X(u)$ does not give an analytical form for  $R(u)$, it is better to use an approximated one, depending on the quantum parameter $\zeta = \omega G_0 \xi_0/r_s^2 \simeq \omega l_p^2/r_s^2 $. The approximated solution can be obtained for two regions:

\begin{itemize}
\item If $u \gg l_p/r_s$, then $u \gg \sqrt{\zeta}$, and thus, the quantum effects are negligible and we would have $X(u)\equiv X_{>}(u) \simeq A(u)$. For this solution the coordinate $R(u)$ becomes
\begin{equation}
R(u) \equiv R_{>}(u) \simeq u +\ln\abs{u-1} \ .
\end{equation}

\item If $u \ll l_p/r_s$, then $u \ll \sqrt{\zeta}$, and thus 
\begin{equation}
\frac{X'(u)}{X(u)} \simeq \frac{u^2}{\zeta} \frac{A'(u)}{A(u)} \ ,
\end{equation}
so that for this region
\begin{equation}
X(u) \equiv X_{<}(u) = X_0 \exp(\frac{1}{\zeta} (u+\ln{\abs{u-1}}))
\end{equation}
where the constant $X_0$ is determined by the matching the two solutions at $u\simeq \sqrt{\zeta}$.

As for this second region the solution depends on $\zeta$, one can classify the results in three cases:
\begin{itemize}
\item[\textbf{a)}] \textbf{More--Classical solution.}
If the quantum effects are small ($ \zeta \ll 1 $), then for the region $u \ll \sqrt{\zeta}$, we can approximate $X_{<}(u) \simeq X^{\text{(MC)}}_{<}(u)$ as
\begin{equation} \label{X-MC}
X^{\text{(MC)}}_{<}(u) = X_0 (1 -\frac{u^2}{2\zeta}) \ .
\end{equation}
The identification $X^{\text{(MC)}}_{<}(\sqrt{\zeta})=X_{>}(\sqrt{\zeta})$, leads to $X_0 = 2(\sqrt{\zeta}-1)/\sqrt{\zeta}$, and we would have
\begin{equation}
R^{\text{(MC)}}_{<} (u) = \frac{1}{X_0} (u+\frac{u^3}{6\zeta}) + C
\end{equation}
where $C$ is a constant. After using the compactification relations (\ref{comp1})-(\ref{comp2}), the Penrose--Carter diagram of this space--time, can be obtained as figure (\ref{MC-P}).

The solution \eqref{X-MC} is restricted to the region where the conditions $\zeta \ll 1$ and $u \ll \sqrt{\zeta}$ are satisfied. Therefore, the quantum effects are not present outside the horizon, where $u>1$. On the other hand, these conditions are well satisfied around the singularity. It can be seen that the improved singularity is fared away from the classical one, leading to the compactification of the $T$-constant surfaces.

\begin{figure}
    \centering
		\includegraphics[width=0.7\textwidth]{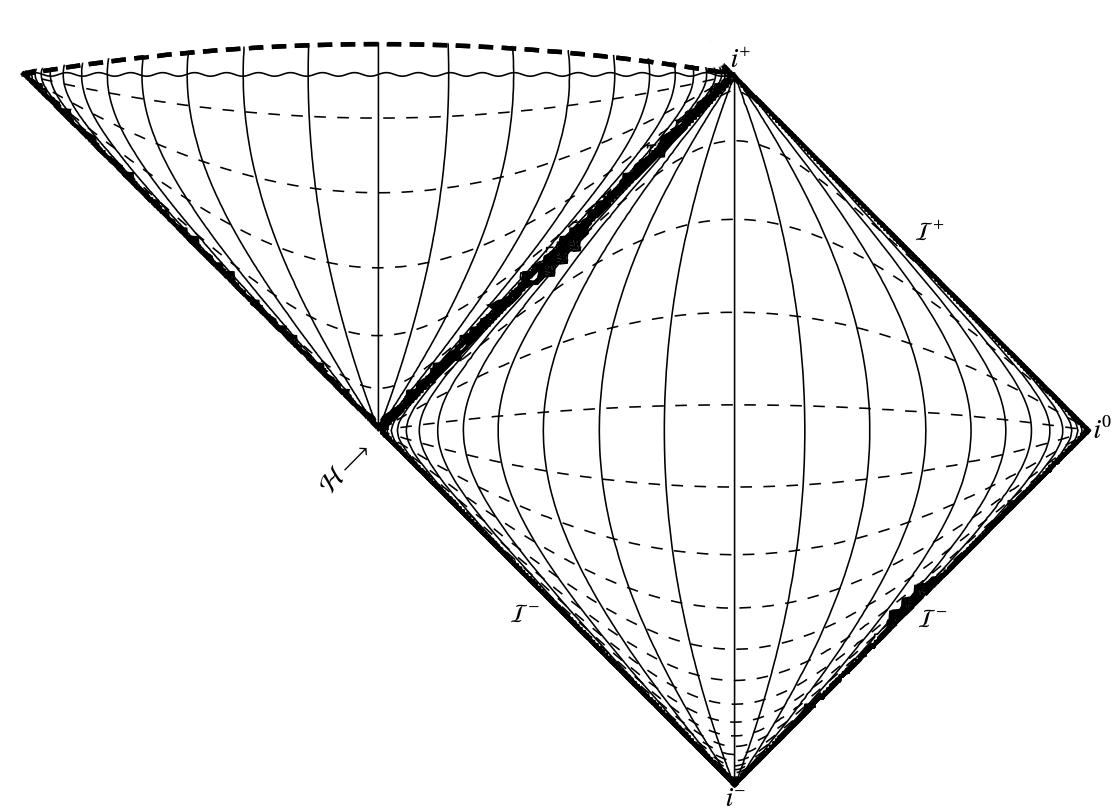}
	\caption{The Penrose--Carter diagram for the More--Classical solution with $\zeta=0.1$. The dashed lines describe $T$-constant surfaces while the thin ones are $R$-constant surfaces. The wavy line is the classical singularity, and the thick dashed line is the improved singularity.}
	\label{MC-P}
\end{figure}

\item[\textbf{b)}] \textbf{More--Quantum solution.}
If the quantum effects are described by $ \zeta \sim 1 $ (i.e. when the Schwarzschild radius is about the Planck length, $r_s \sim l_p$), then we can approximate $X_{<}(u) \simeq X^{\text{(MQ)}}_{<} (u)$ by
\begin{equation}
X^{\text{(MQ)}}_{<}(u) = X_0 e^{-u^2/2\zeta} \ .
\end{equation}
After determining $X_0$ by the relation $X^{\text{(MQ)}}_{<}(\sqrt{\zeta})=X_{>}(\sqrt{\zeta})$, we can find the $R^{\text{(MQ)}}_{<}(u)$,
\begin{equation}
 R^{\text{(MQ)}}_{<}(u) = \frac{\zeta}{\sqrt{e}(\sqrt{\zeta}-1)} \sqrt{\frac{\pi}{2}} \text{Erfi}(\frac{u}{\sqrt{2\zeta}}) + C
\end{equation}
where $C$ is a constant. The Penrose--Carter diagram of this space--time, can be seen in figure (\ref{MQ-P}).

In this case, the quantum antiscreening effects can be observed more around the horizon, since $\zeta\sim 1$. Towards the singularity, the clock of an observer lying on the $T$-constant surfaces ticks faster because of the anti gravity effects of the antiscreening coupling. Remember that the choice of $f=\zeta / u^2$ leads to strong quantum effects inside the horizon ($u < 1$).

\begin{figure}
    \centering
		\includegraphics[width=0.7\textwidth]{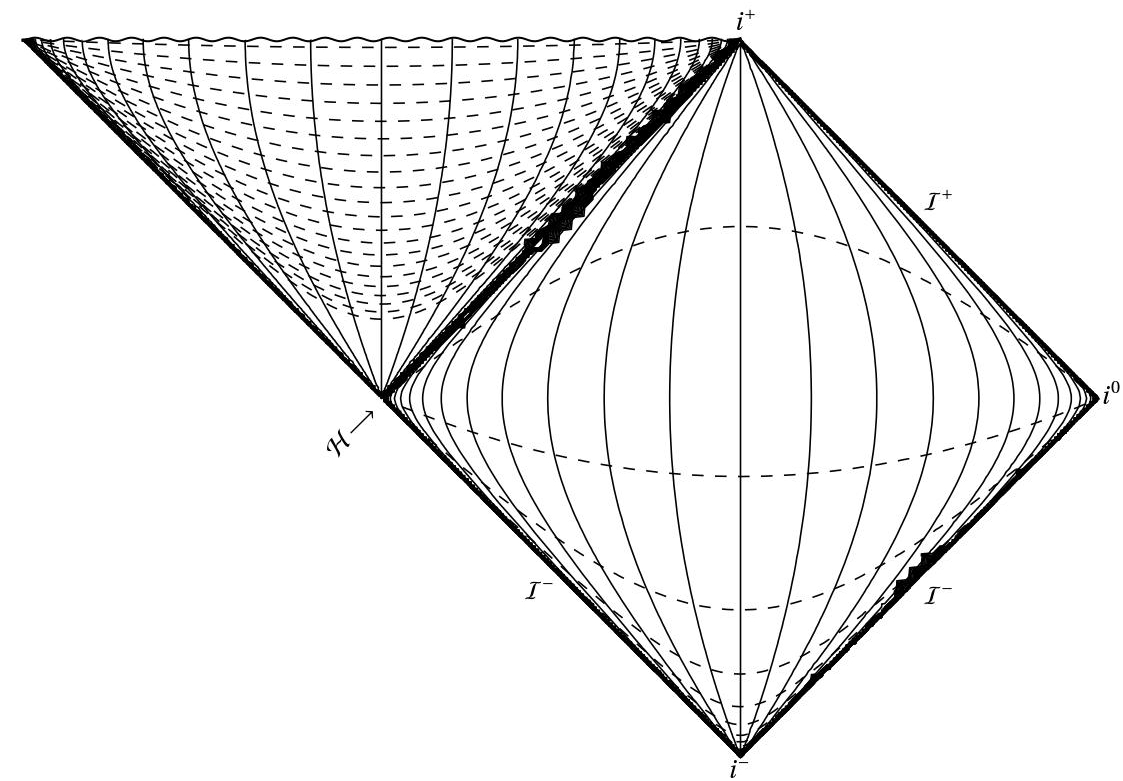}
	\caption{The Penrose--Carter diagram for the More--Quantum solution with $\zeta=1.1$. The dashed lines describe $T$-constant surfaces while the thin ones are $R$-constant surfaces. The wavy line is the singularity.}
	\label{MQ-P}
\end{figure}

\item[\textbf{c)}] \textbf{Pure--Quantum solution.} Finally, if the quantum effects are large enough such that $ \zeta \gg 1 $ (where the Schwarzschild radius is much smaller than Planck length, $r_s \ll l_p$), then we can approximate $X_{<}(u) \simeq X^{\text{(PQ)}}_{<}$ by
\begin{equation}
X^{\text{(PQ)}}_{<}(u) = X_0 e^{-u^2/2\zeta+u^3/3\zeta} \simeq (1-\frac{u^2}{2\zeta})(1+\frac{u^3}{3\zeta})
\end{equation}
where $X_0= (1-1/\sqrt{\zeta})\exp(1/2-\sqrt{\zeta}/3)$. Then,
\begin{equation}
 R^{\text{(PQ)}}_{<}(u) = -\frac{u^6/6 + u^3\zeta-u^4\zeta/2-6u\zeta^2}{6\zeta^2} + C
\end{equation}
where $C$ is a constant. 

Since,$u \ll \zeta$ and $\zeta \gg 1$ for this solution, the main compactification occurs far from the singularity and near the horizon, figure (\ref{TQ-P}). Like for the More--Quantum case, because of $f=\zeta / u^2$ the main effects are present inside horizon.

\begin{figure}
    \centering
		\includegraphics[width=0.7\textwidth]{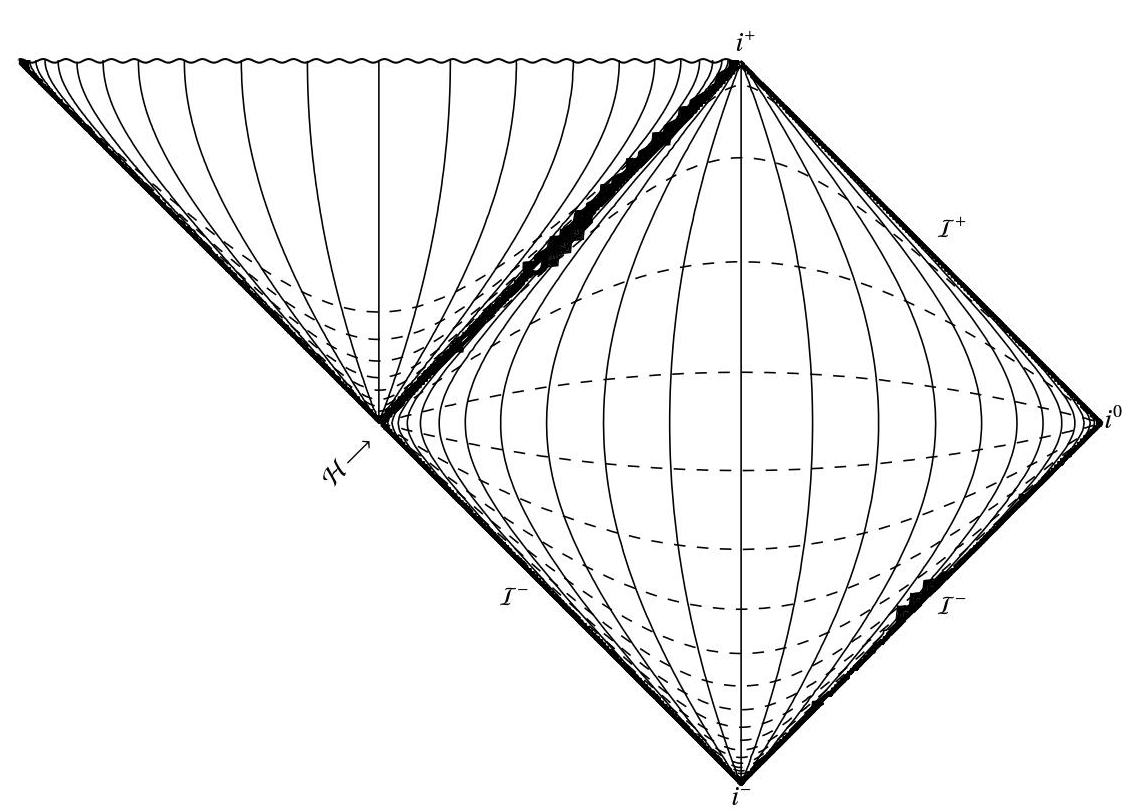}
	\caption{The Penrose--Carter diagram for the Pure--Quantum solution with $\zeta=1.5$. The dashed lines describe $T$-constant surfaces while the thin ones are $R$-constant surfaces. The wavy line is the singularity.}
	\label{TQ-P}
\end{figure}

\end{itemize}

\end{itemize}

\section{Connection as a dynamical object}
The results of the previous section shows that quantum improvement of the affine connection can leads to non--trivial results. In order to have a complete model for quantum improvement of the affine, we have to do it within a pure affine theory of gravity.
To construct a gravity theory which considers the affine connection as a dynamical object, we need to define a  covariant action in terms of affine. But as the connection does not transform covariantly, it cannot be used  by itself to define a proper action. 
There are diverse combinations of the connection and its canonical momentum for this purpose. 
To this end, the Riemann curvature $R^{\lambda}_{\mu\nu\kappa}$ (as the canonical momentum) is a good candidate. Besides its presence in the description of the tidal forces as the gravitational force, it is quiet useful for our purpose. This is because we have to improve the action using the curvature invariants $\chi_i$.  Schr\"{o}dinger--Eddington action provides such a model and thus in what follows we present an improved Schr\"{o}dinger--Eddington action.

In order to bring quantum modifications into such an affine theory, one should use the running gravitation coupling obtained from the effective average action of an affine theory. But it should be brought in mind that we have stuck to Einstein--Hilbert truncation (in the effective action theory space), and this 2-dimensional subspace has no coupling to torsion and non-metricity. Therefore, it seems that there is no serious deviations in the results of the effective average action theory based on the metric theory and the one based on the affine theory. This allows  to use the running coupling constant obtained from an Einstein--Hilbert truncated metric theory for our purpose.

\subsection{Improved Schr\"{o}dinger--Eddington action}
Historically, Schr\"{o}dinger used the contracted Riemann tensor $R^{\nu}_{\mu\nu\kappa}$ , i.e. Ricci tensor   $R_{\mu\kappa}$, as a covariant tensor of rank two to construct the measure of integration for the curved space--time\cite{Schrodinger-B}. This choice was based on the Eddington, Einstein  and Straus endeavor to unify the gravitation and electromagnetism\cite{Eddington-B,E-S}.

Keeping in mind that the metric is not a dynamical object in the pure affine gravity, the generalized Sch\"{o}dinger--Eddington action is
\begin{equation} \label{ES-A}
\mathcal{A}_{\text{SE}} = \int \dd[4]{x} \mathcal{L}[\Psi^A,\Psi^{A}_{\ ;\kappa}]\sqrt{-\det R_{\mu\nu}} \ ,
\end{equation}
where $\Psi^A$ stands for a general matter field of any rank. The covariant derivative of this field could be defined as
\begin{equation}
\Psi^{A}_{\ ;\kappa} = \Psi^{A}_{\ ,\kappa} +C^{A \ \nu}_{\ B \ \mu} \ \Gamma^{\mu}_{\nu\kappa} \ \Psi^{B}
\end{equation}
where the coefficient $C^{A \ \nu}_{\ B \ \mu}$ depends on the nature of the matter fields. 
The action is a functional of $\Gamma^\alpha_{\beta\gamma}$ as the dynamical variable, and $R_{\alpha\beta}$ as the canonical momentum. 
The metric can be constructed as $g^{\alpha\beta} \equiv \text{g}^{\alpha\beta} / \sqrt{-\det g^{\alpha\beta}}$ with the metric density defined as
\begin{equation} \label{MD}
\text{g}^{\alpha\beta} =  \pdv{\mathcal{L}\sqrt{-\det R_{\mu\nu}}}{R_{\alpha\beta}} \ .
\end{equation}

We use the action \eqref{ES-A} to construct the quantum improved affine theory. To improve the coupling constant, the implicit dependence of the Lagrangian density $ \mathcal{L}[\Psi^A,\Psi^{A}_{\ ;\kappa}]$ on the coupling should be exprssed explicitly, i.e. the improved action would be
\begin{equation}
\mathcal{A}^{(\mathcal{I})}_{\text{SE}} = \int \dd[4]{x} \mathcal{L}[\Psi^A,\Psi^{A}_{\ ;\kappa},G(\chi)]\sqrt{-\det R_{\mu\nu}}  \ .
\end{equation}

We have to note that there is a mathematical debate on the possibility of construction of proper Lagrangian for various matter fields without definition of any fundamental metric field. Although this is not important in what follows (since we will concentrate on vacuum solution), but in fact inverse of Ricci tensor can be used for this purpose. The validity of this suggestion can be tracked in the identity
\begin{equation}
\delta \sqrt{-\det R_{\mu\nu}} = \frac{1}{2} \sqrt{-\det R_{\mu\nu}}\ K^{\alpha\beta}\ \delta R_{\alpha\beta} \ ,
\end{equation}
which is used to define the Ricci inverse $K^{\alpha\beta} \equiv (R_{\alpha\beta})^{-1} $.
A serious debate on this issue can be found in Ref. \cite{Metric}.

The dependence of the running coupling $G(\chi)$ on the affine connection and its derivatives is through the curvature invariants. As it is previousely discussed (\cite{our4,our5}), although to have an exact scaling interpretation, considering all the sixteen curvature invariants is expected, but some conditions such as singularities and energy conditions restrict our choices\cite{our4,our5}. For simplicity of calculations, we just use the essential invariants ($\chi_1\equiv R, \ \chi_2\equiv R_{\mu\nu\kappa\lambda}R^{\mu\nu\kappa\lambda}, \ \chi_3\equiv R_{\mu\nu}R^{\mu\nu}$) for the scaling process, but the method could simply extended to more general cases.

The field equations are derived from the least action principle as usual, and thus we need to evaluate
\begin{equation}
\delta\mathcal{A}^{(\mathcal{I})}_{\text{ES}} = \int  \dd[4]{x}  \delta\left(\mathcal{L}[\Psi^A,\Psi^{A}_{\ ;\kappa},G(\chi)]\sqrt{-\det R_{\mu\nu}} \right)\ .
\end{equation}

Straightforward calculations end to
\begin{equation}
\delta\mathcal{A}^{(\mathcal{I})}_{\text{SE}} = \delta\mathcal{A}^{(\mathcal{I})}_{\text{G}} +  \delta\mathcal{A}^{(\mathcal{I})}_{\text{M}}
\end{equation}
where each term is defined as
\begin{align} 
&  \delta\mathcal{A}^{(\mathcal{I})}_{\text{G}} = \int  \dd[4]{x} \left( \sqrt{-\det R_{\mu\nu}}\pdv{\mathcal{L}}{\Gamma^{\gamma}_{\alpha\beta}} \delta\Gamma^{\gamma}_{\alpha\beta} +  \sqrt{-\det R_{\mu\nu}}\pdv{\mathcal{L}}{\Gamma^{\gamma}_{\alpha\beta;\kappa}} \delta\Gamma^{\gamma}_{\alpha\beta;\kappa} + \pdv{\mathcal{L}\sqrt{-\det R_{\mu\nu}}}{R_{\alpha\beta}} \delta R_{\alpha\beta} \right) \ , \\
\label{PAAV}
&  \delta\mathcal{A}^{(\mathcal{I})}_{\text{M}} = \int  \dd[4]{x} \left(\pdv{  \sqrt{-\det R_{\mu\nu}} \mathcal{L}}{\Psi^{B}} \delta\Psi^{B} +  \pdv{\sqrt{-\det R_{\mu\nu}}\mathcal{L}}{ \Psi^{B}_{,k}} \delta\Psi^{B}_{,k} \right)
\end{align}
and they should vanish individually, and we have used the notation $\mathcal{L}\equiv \mathcal{L}[\Psi^A,\Psi^{A}_{\ ;\kappa},G(\chi)]$.

There is a clear concern on the dependence of the running coupling $G(\chi)$ on the Ricci square invariant. To solve this concern, we rearrange the scaling parameter dependence as $\chi \equiv \chi(\Gamma^{\gamma}_{\alpha\beta}, \Gamma^{\gamma}_{\alpha\beta;\kappa}, R_{\alpha\beta})$.
Thus \eqref{PAAV} leads to
\begin{multline} \label{V-ES}
\delta\mathcal{A}^{(\mathcal{I})}_{\text{G}} = \int  \dd[4]{x} \sqrt{-\det R_{\mu\nu}}  \left(  \mathcal{M}_A^{\ \kappa}\pdv{\Psi^A_{\ ;\kappa}}{\Gamma^{\gamma}_{\alpha\beta}} \ \delta\Gamma^{\gamma}_{\alpha\beta} + 
\sum_{i=1}^{2} \mathcal{L}_{\chi^{(i)}} ( \mathcal{X}_{\gamma}^{\ \alpha\beta  \ (i)} \ \delta\Gamma^{\gamma}_{\alpha\beta} +  \mathcal{N}_{\gamma}^{\ \alpha\beta\kappa  \ (i)} \ \delta\Gamma^{\gamma}_{\alpha\beta;\kappa} ) \right) \\
+ \int  \dd[4]{x} \left( \pdv{\mathcal{L}\sqrt{-\det R_{\mu\nu}}}{R_{\alpha\beta}} \ \delta\Gamma^{\gamma}_{\alpha\beta;\gamma} - \pdv{\mathcal{L}\sqrt{-\det R_{\mu\nu}}}{R_{\alpha\beta}} \ \delta\Gamma^{\gamma}_{\alpha\gamma;\beta} 
- 2 \pdv{\mathcal{L}\sqrt{-\det R_{\mu\nu}}}{R_{\alpha\kappa}} S^{\beta}_{\gamma\kappa}  \ \delta\Gamma^{\gamma}_{\alpha\beta} \right)
\end{multline}
where
\begin{align}
& \mathcal{L}_{\chi_{(i)}} =  \pdv{\mathcal{L}}{G(\chi)} \pdv{G(\chi)}{\chi^{(i)}} \ ,\\
& \mathcal{X}_{\gamma}^{\ \alpha\beta  \ (i)} = \pdv{\chi_{(i)}}{\Gamma^{\gamma}_{\alpha\beta \ (i)}} \ , \\
& \mathcal{N}_{\gamma}^{\ \alpha\beta\kappa  \ (i)} =  \pdv{\chi_{(i)}}{\Gamma^{\gamma}_{\alpha\beta;\kappa \ (i)}} \ , \\
&  \mathcal{M}_A^{\ \kappa} = \pdv{\mathcal{L}}{\Psi^A_{\ ;\kappa}} 
\end{align}
and the Cartan torsion tensor $S^{\kappa}_{\mu\nu} \equiv \Gamma^{\kappa}_{[\mu\nu]}$ is the antisymmetric part of the connection.

The field equations lead to a differential equation between the constructed metric $g^{\alpha\beta}$ and the affine connection $\Gamma^{\sigma}_{\alpha\beta}$. Using the Gauss theorem, we get  
\begin{equation}
\text{g}^{\alpha\beta}_{\ ;\gamma} - \text{g}^{\alpha\kappa}_{\ ;\kappa} \delta^{\beta}_{\gamma} - 2 \text{g}^{\alpha\beta} S_{\gamma} +2 \text{g}^{\alpha\kappa}   S_{\kappa}\delta^{\beta}_{\gamma} +2 \text{g}^{\alpha\kappa} S^{\beta}_{\gamma\kappa}
= \mathcal{S}^{\alpha\beta \ (M)}_{\ \gamma} + \mathcal{S}^{\alpha\beta \ (Q)}_{ \ \gamma}
\end{equation}
as the quantum improved field equations,
where  $ S_{\mu} \equiv S^{\nu}_{\ \mu\nu}$ is the torsion vector and
\begin{align}
& \mathcal{S}^{\alpha\beta \ (M)}_{\ \gamma} = \sqrt{-\det R_{\mu\nu}} \mathcal{M}_A^{\ \beta} \ C^{A \ \alpha}_{\ B \ \gamma} \ \Psi^B  \ , \\
&  \mathcal{S}^{\alpha\beta \ (Q)}_{ \ \gamma} =
\sqrt{-\det R_{\mu\nu}}  \sum_{i=1}^{2} \mathcal{L}_{\chi^{(i)}} \left( \mathcal{X}_{\gamma}^{\ \alpha\beta  \ (i)}  - \frac{2}{3} \mathcal{N}_{\gamma}^{\ \alpha\beta\kappa  \ (i)} S_{\kappa} - \nabla_{\kappa} \mathcal{N}_{\gamma}^{\ \alpha\beta\kappa  \ (i)}  \right)  
\end{align}
are matter and quantum improvement source terms.
By contracting $(\alpha,\gamma)$ the generalized equation of Schr\"{o}dinger theory would be derived,
\begin{equation} \label{IPAFE-1}
\text{g}^{[\alpha\beta]}_{\ ,\alpha}= \delta^{\gamma}_{\alpha}  \mathcal{S}^{\alpha\beta \ (M)}_{\ \gamma} + \delta^{\gamma}_{\alpha} \mathcal{S}^{\alpha\beta \ (Q)}_{ \ \gamma} \ .
\end{equation}
The inhomogeneous source at the right hand side of the field equations \eqref{IPAFE-1} is constructed from two terms: The matter source $ \delta^{\gamma}_{\alpha}  \mathcal{S}^{\alpha\beta \ (M)}_{\ \gamma}$ and the pure quantum part $ \delta^{\gamma}_{\alpha} \mathcal{S}^{\alpha\beta \ (Q)}_{ \ \gamma}$. The  $ \delta^{\gamma}_{\alpha}  \mathcal{S}^{\alpha\beta \ (M)}_{\ \gamma}$ does not include dynamical effects of the quantum correction of the improvement while the prescribed running coupling $G(\chi)$ attends in this term. On the other hand, the pure quantum term $\delta^{\gamma}_{\alpha} \mathcal{S}^{\alpha\beta \ (Q)}_{ \ \gamma}$ contains both dynamical and non--dynamical effects arising from quantum improvement.

One can rewrite the field equations \eqref{IPAFE-1} as differential equations for the metric tensor $g_{\alpha\beta}$. From \eqref{V-ES} and \eqref{MD} we would have
\begin{equation} \label{IPAFE-2}
g^{\alpha\beta}_{\ ;\gamma} \sqrt{-\det g_{\mu\nu}} - g^{\alpha\beta} (\sqrt{-\det g_{\mu\nu}})_{\ ,\gamma} - \text{g}^{\alpha\beta} \Gamma^{\kappa}_{\kappa\gamma}  - 2 \text{g}^{\alpha\beta} S_{\gamma} 
+2 \text{g}^{\alpha\kappa}   \left(  S^{\beta}_{\gamma\kappa} + \frac{1}{3} S_{\kappa}\delta^{\beta}_{\gamma} \right)
= \mathcal{S}^{\alpha\beta \ (C)}_{\ \gamma} + \mathcal{S}^{\alpha\beta \ (Q)}_{ \ \gamma}
\end{equation} 
Multiplying this evolution equation by the inverse of the metric $g_{\alpha\beta}$, leads to
\begin{equation} \label{Tr-IPAFE-2}
 (\sqrt{-\det g_{\mu\nu}})_{\ ,\gamma} -  \sqrt{-\det g_{\mu\nu}}  \Gamma^{\kappa}_{\kappa\gamma} = \frac{8}{3} \sqrt{-\det g_{\mu\nu}} S_{\gamma} + g_{\alpha\beta} \left( \mathcal{S}^{\alpha\beta \ (C)}_{\ \gamma} + \mathcal{S}^{\alpha\beta \ (Q)}_{ \ \gamma} \right) \ .
\end{equation}
And on substituting \eqref{Tr-IPAFE-2} in \eqref{IPAFE-2} the covariant form
\begin{equation} \label{Co-IPAFE-2}
g^{\alpha\beta}_{\ ;\gamma} + 2   S^{\beta}_{\gamma\kappa} g^{\alpha\kappa}  + \frac{2}{3} S_{\gamma}g^{\alpha\beta} +\frac{2}{3} S_{\kappa}\delta^{\beta}_{\gamma} g^{\alpha\kappa}= -3 \left( \mathcal{S}^{\alpha\beta \ (C)}_{\ \gamma} + \mathcal{S}^{\alpha\beta \ (Q)}_{ \ \gamma} \right)
\end{equation}
is obtained.

Using the Schr\"{o}dinger star--affine connection
\begin{equation} \label{SA}
\prescript{*}{}\Gamma^{\gamma}_{\alpha\beta} = \Gamma^{\gamma}_{\alpha\beta} + \frac{2}{3}\delta^{\gamma}_{\alpha} S_{\beta}
\end{equation}
the covariant equation \eqref{Co-IPAFE-2} would have the more familiar form
\begin{equation} \label{Co-IPAFE-3}
g_{\alpha\beta,\gamma} - \prescript{*}{}\Gamma^{\kappa}_{\alpha\gamma} g_{\kappa\beta} - \prescript{*}{}\Gamma^{\kappa}_{\gamma\beta} g_{\alpha\gamma} = -3 g_{\alpha\mu} g_{\beta\nu} \left( \mathcal{S}^{\mu\nu \ (C)}_{\ \gamma} + \mathcal{S}^{\mu\nu \ (Q)}_{ \ \gamma} \right) \ .
\end{equation}
Same as its classical counterpart\cite{Schrodinger-B}, 64 ordinary linear equations for the star affine connection are derived from equation \eqref{Co-IPAFE-3}. To continue, we need 16 components of the metric tensor $g_{\alpha\beta}$ and 4 components of torsion vector $S_{\alpha}$.  Thus, we can use the definition of metric density \eqref{MD} and the star--affine \eqref{SA} to obtain
\begin{equation} \label{AS-MD}
\prescript{*}{}R_{\alpha\beta}+ \frac{2}{3} (S_{\alpha,\beta}-S_{\beta,\alpha})= \frac{2}{\mathcal{L}} \left(g_{\alpha\beta}-\pdv{\mathcal{L}}{\prescript{*}{}R^{\mu\nu}}\delta^{\mu}_{\alpha}\delta^{\nu}_{\beta} \right) \ .
\end{equation}
To sum up, the set of coupled equations \eqref{IPAFE-1}, \eqref{Co-IPAFE-3} and \eqref{AS-MD} are 84 equations for 84 components of the star--affine connection $ \prescript{*}{}\Gamma^{\gamma}_{\alpha\beta}$, torsion vector $S_{\alpha} $ and the metric field $g_{\alpha\beta} $.

\subsection{Spherically symmetric vacuum solution}
Since the static spherically symmetric space--time is a notable solution in any gravitational theory, we look for such a solution for our quantum improved affine model. Therefore, using this symmetry, we find a solution of equations \eqref{IPAFE-1}, \eqref{Co-IPAFE-3} and \eqref{AS-MD} for a vacuum Lagrangian $\mathcal{L} \equiv 2/\Lambda G(\chi)$.

The general metric tensor for this space--time is the one that its functional form is not changed by the rotation about the center of the symmetry. Papapetrou suggested\cite{Papapetrou} the general form as 
\begin{equation} \label{SSM}
 g_{\alpha\beta} = 
\begin{pmatrix}
\gamma & -\omega & 0 & 0 \\
\omega & -\alpha & 0 & 0 \\
0 & 0 & -\beta & r^2 v \sin\theta \\
0 & 0 & -r^2 v \sin\theta & -\beta \sin^2\theta
\end{pmatrix}
\end{equation}
which satisfies the same symmetry in the Einstein--Straus Theory. The functions $\alpha\equiv\alpha(r)$, $\beta\equiv\beta(r)$, $\gamma\equiv\gamma(r)$, $\omega\equiv\omega(r)$ and $v\equiv v(r)$ should be determined from the equations \eqref{IPAFE-1}, \eqref{Co-IPAFE-3} and \eqref{AS-MD}. Clearly, the choice $\beta=r^2$ does not ruins any generality. Indeed, although the both cases ($\omega\neq 0,v= 0$) and ($\omega=0,v\neq 0$) were studied by Papapetrou as solutions to the Einstein--Straus theory\cite{Papapetrou}, we examine just the first one, since the improved \textit{metric} gravity of this solution has been studied widely\cite{Bonanno & Reuter,our2,Pawlowski} and make the comparison of the results possible.

For the same reasons, we choose the scaling parameter $\chi=R_{\alpha\beta}R^{\alpha\beta}$. While the quantum effects at the right hand side of the equations \eqref{IPAFE-1} and  \eqref{Co-IPAFE-3} are removed by this cut off identification, 
these effects can be followed from \eqref{AS-MD}. 
Since the quantities $ \mathcal{X}_{\gamma}^{\ \alpha\beta  \ (i)} $, $ \mathcal{N}_{\gamma}^{\ \alpha\beta\kappa  \ (i)} $ and $  \mathcal{M}_A^{\ \kappa} $ vanish for our chosen cutoff identification  $\chi=R_{\alpha\beta}R^{\alpha\beta}$, the inhomogeneous terms $\mathcal{S}^{\mu\nu \ (C)}_{\ \gamma} $ and $ \mathcal{S}^{\mu\nu \ (Q)}_{ \ \gamma}$ at the right hand side of the equations \eqref{IPAFE-1} and \eqref{Co-IPAFE-3} disappear. Therefore, like the classical case, from \eqref{IPAFE-1} we would have
\begin{equation} \label{l}
\frac{\omega^2 r^4}{\alpha \gamma -\omega^2} = l^4
\end{equation}
where $l$ is a constant of dimension of length.

The equation \eqref{Co-IPAFE-3} with zero right hand side, is a linear algebraic equation for the connection components. The non--vanishing star affine components are then \cite{EM-GR1,Papapetrou}
\begin{align}
& \prescript{*}{}\Gamma^{t}_{tr} = \prescript{*}{}\Gamma^{t}_{rt}  = \frac{\gamma'}{2\gamma} + \frac{2\omega^2}{r\alpha\gamma} \ , \nonumber \\
& \prescript{*}{}\Gamma^{r}_{tt} = \frac{\gamma'}{2\alpha}+\frac{4\omega^2}{r\alpha^2} , \quad \prescript{*}{}\Gamma^{r}_{rr}=\frac{\alpha'}{\alpha} , \quad \prescript{*}{}\Gamma^{r}_{\theta\theta}=-\frac{r}{\alpha} , \quad \prescript{*}{}\Gamma^{r}_{\varphi\varphi} = -\frac{r}{\alpha}\sin^2\theta \ ,\nonumber  \\
&\prescript{*}{}\Gamma^{\theta}_{r\theta}=\prescript{*}{}\Gamma^{\theta}_{\theta r}=\frac{1}{r} , \quad \prescript{*}{}\Gamma^{\theta}_{\varphi\varphi}=-\sin\theta\cos\theta  \ , \nonumber \\
&\prescript{*}{}\Gamma^{\varphi}_{r\varphi}=\prescript{*}{}\Gamma^{\varphi}_{\varphi r}=\frac{1}{r} , \quad \prescript{*}{}\Gamma^{\varphi}_{\theta\varphi}=\prescript{*}{}\Gamma^{\varphi}_{\varphi\theta}=\cot\theta  \ , \nonumber \\
&\prescript{*}{}\Gamma^{r}_{tr}=-\prescript{*}{}\Gamma^{r}_{rt}=-\frac{2\omega}{r\alpha} , \quad \prescript{*}{}\Gamma^{\theta}_{t\theta}=-\prescript{*}{}\Gamma^{\theta}_{\theta t}=\prescript{*}{}\Gamma^{\varphi}_{t\varphi}=-\prescript{*}{}\Gamma^{\varphi}_{\varphi t}= \frac{\omega}{r\alpha}
\end{align}
where the prime denotes derivative with respect to $r$.

At last, we can use equation  \eqref{AS-MD} to get the Ricci tensor and torsion vector components. As the constant $\xi$ contains the quantum effects which are small, we can consider the quantum terms in equation \eqref{AS-MD} as a perturbation to the classical solution.

The Ricci tensor components would be
\begin{align}
& \prescript{*}{}R_{tt}  = \left(\frac{\gamma'}{2\alpha}+\frac{4\omega^2}{r\alpha^2} \right)'- \left(\frac{\gamma'}{2\alpha}+\frac{4\omega^2}{r\alpha^2} \right) \left(\frac{\gamma'}{2\gamma}+\frac{2\omega^2}{r\alpha\gamma}-\frac{\alpha'}{\alpha}+\frac{2}{r} \right) + \frac{6\omega^2}{r^2\alpha^2}  = -\frac{\Lambda \gamma}{1-f}  \ , \label{R00} \\
& \prescript{*}{}R_{rr} = \frac{\alpha'}{r\alpha}-\left(\frac{\gamma'}{2\gamma}+\frac{2\omega^2}{r\alpha\gamma}\right)'+\left(\frac{\gamma'}{2\gamma}+\frac{2\omega^2}{r\alpha\gamma}\right)\left(\frac{\alpha'}{2\alpha}-\frac{\gamma'}{2\gamma}-\frac{2\omega^2}{r\alpha\gamma}\right) = \frac{\Lambda \alpha}{1-f}  \ , \label{R11} \\
& \prescript{*}{}R_{\theta\theta}  = \frac{1}{\sin^2\theta} \prescript{*}{}R_{\varphi\varphi} =  1-\frac{1}{\alpha} + \frac{r\alpha'}{2\alpha^2} -\frac{r}{\alpha} \left(\frac{\gamma'}{2\gamma}+\frac{2\omega^2}{r\alpha\gamma} \right) \frac{\Lambda r^2}{1-f} \ , \label{R22} \\
& \prescript{*}{}R_{tr} = - \prescript{*}{}R_{rt} = -\left(\frac{2\omega}{r\alpha}\right)'-\frac{4\omega}{r^2\alpha}  =\frac{\Lambda\omega}{1-f}-\frac{2}{3} S'_0  \label{R01}
\end{align}
where for small $\Lambda\ll 1$, and up to the first order
\begin{footnotesize}
\begin{multline}
f\equiv f(R_{\alpha\beta}R^{{\alpha\beta}}) =  \frac{2\xi}{3} \times \\ \sqrt{\frac{
3l^4r^4(24r r_s-18r_s^2-r^3\Lambda(4r+9 r_s))+3r^8r_s(3r_s-r^3\Lambda)+l^8(162r^2-288 r r_s +153 r_s^2-6\Lambda r^3(  22 r-14 r_s ))}{r^{14}}}
\end{multline}
\end{footnotesize}
Thus a set of solutions for $\gamma$, $\alpha$, $\omega$ and  $S_0$ are derived from equations \eqref{R00}-\eqref{R01} and \eqref{l}.

The classical solutions for $\gamma$, $\alpha$ and $\omega$ (i.e. ignoring quantum terms) are
\begin{align}
& \gamma_{Class.} \equiv \bar{\gamma} = (1-\frac{r_s}{r}-\frac{\Lambda r^2}{3})(1+\frac{\bar{l}^4}{r^4}) \ , \\
& \alpha_{Class.} \equiv \bar{\alpha} =  (1-\frac{r_s}{r}-\frac{\Lambda r^2}{3})^{-1} \ , \\
& \omega_{Class.} \equiv \bar{\omega} =  \pm\frac{\bar{l}^2}{r^2}
\end{align}
where $\bar{l}^4 = \bar{\omega}^2 r^4 / (\bar{\alpha}\bar{\gamma}-\bar{\omega}^2)$ is a length dimensional constant of integration.
Writing the quantum solution as a perturbation added to the classical one
\begin{align}
& \gamma = \bar{\gamma} + \phi_1 \ , \\
& \alpha = \bar{\alpha} + \phi_2 \ , \\
& \omega = \bar{\omega} + \phi_3 \ ,
\end{align} 
and for small $u$, where the quantum effects are important, from \eqref{R11}-\eqref{R22} we get the coupled differential equations
\begin{equation}
\frac{(u-1)^2}{2u}\dot{\phi}_2(u) + \frac{(u-1)}{u}\phi_2(u) + \frac{4u}{l^2_s} \phi_3(u) = 
- \frac{u^2(-17+2\sqrt{17}\sqrt{\frac{l^8_s}{u^{14}}}(17+16 u^2)\zeta)\Lambda_s}{17(1-2\sqrt{17}\sqrt{\frac{l^8_s}{u^{14}}}\zeta)^2} \ ,
 \, \label{CDE-1}
\end{equation}
\begin{equation}
\ddot{\phi}_2(u)- \frac{8}{2u}\dot{\phi}_2(u) + \frac{31}{2u^2}\phi_2(u)  = - \frac{2(187+3u(17-16u(3+u)))}{\sqrt{17}u^5}l^4_s\Lambda_s\zeta   \label{CDE-2}
\end{equation}
for $\phi_2$ and $\phi_3$. Note that $u\equiv r/r_s$, $\Lambda_s \equiv \Lambda r_s^2$, $\zeta \equiv \xi/r_s^2$ and $l_s \equiv l/r_s$ are dimensionless quantities and $r_s$ is Schwarzschild radius.

For $l_s=0$ the Schwarzschild solution of the metric theory is rederived. 
It can be shown that the
\begin{align}
& \phi_2(u) \simeq \frac{l^4_s \Lambda_s}{u^3}  \Bigl( 0.881 + 0.165 u \Bigr)\zeta   \\
& \phi_3(u) \simeq \frac{l^6_s \Lambda^2_s}{u^6} \Bigl( 0.14 +0.72 u \Bigr)\zeta
\end{align}
satisfy the couple differential equations \eqref{CDE-1} and \eqref{CDE-2}. 

On the other hand, equation \eqref{l} leads to
\begin{equation}
-l^4_s+\bar{l}^4_s \left(1+ \frac{2\phi_3}{\omega} - \bar{l}^4_s \frac{\alpha\phi_1+\gamma\phi_2-2\omega\phi_3}{\omega^2 r^4} \right) = 0 \ ,  
\end{equation}
and thus
\begin{equation}
\phi_1(u) \simeq \sqrt{\frac{l^8_s}{u^{14}}} l^4_s \Lambda_s \left( \frac{1.05}{u^2} -\frac{1.24}{u}  \right)\zeta
\end{equation}
where the initial condition $l_{s(\zeta=0)} = \bar{l}_s$ is considered.

Afterwards, the spherical static vacuum solution
\begin{align}
& \gamma(u\ll1) = (1-\frac{1}{u}-\frac{\Lambda_s u^2}{3})(1+\frac{l^4_s}{u^4}) + \phi_1(u) \label{gamma} \\
& \alpha(u\ll1) =  (1-\frac{1}{u}-\frac{\Lambda_s u^2}{3})^{-1}+ \phi_2(u) \label{alpha} \\
& \omega(u\ll1) = \pm \left( \frac{l^2_s}{u^2} + \phi_3(u) \right) \label{omega}
\end{align}
would satisfy the set of coupled equations \eqref{IPAFE-1}, \eqref{Co-IPAFE-3} and \eqref{AS-MD}.
Indeed, the only non--zero torsion component for this space--time is
\begin{equation}
r_s S_0(u\ll1) = \pm   \left( l^2_s (\frac{1}{u^3} - \frac{3}{2u^2}-\frac{\Lambda_s}{2u}) + \phi_s(u) \right)
\end{equation}
where
\begin{equation} 
\phi_s(u) =  \frac{l_s^6 \Lambda_s}{u^{10}}  \left(-5.28 + 13.73 u \right) \zeta \ .
\end{equation}
For $\zeta=0$ the classical $S_0$ ($\phi_s =0$) is obtained.

\begin{figure}
    \centering
		\includegraphics[width=0.7\textwidth]{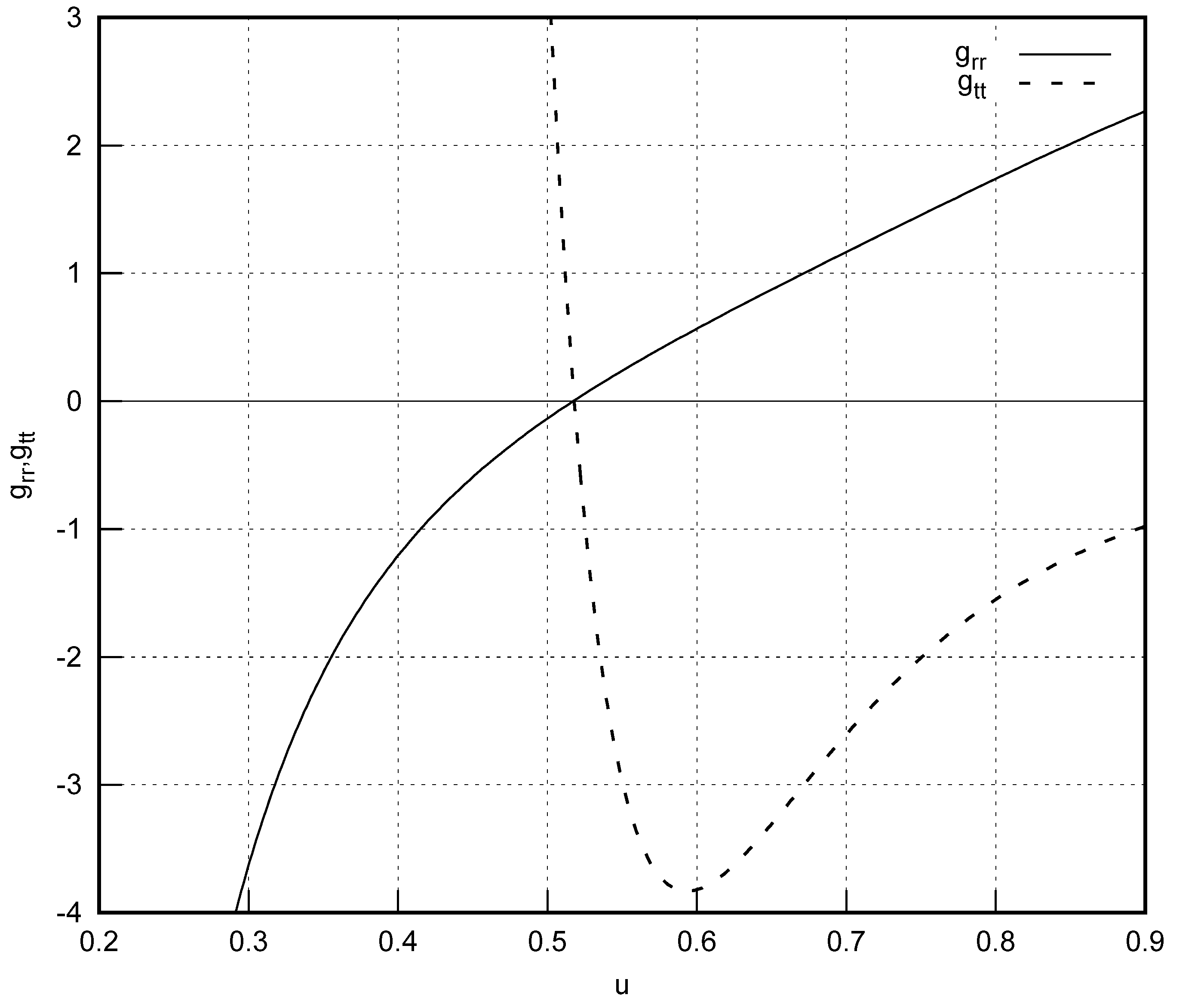}
	\caption{The radial and temporal components of the metric for improved affine gravity with $\zeta = 0.1$.}
	\label{gtt-0.1}
\end{figure}

Let us now, see how the quantum corrections affects the metric components $g_{tt} \equiv \gamma$ and $ g_{rr} \equiv -\alpha$. As it can be seen in  figure \ref{gtt-0.1}, the horizon is shifted. Also near the singularity the metric behavior is dramatically changed.

\section{Conclusions}
Here, we have investigated the effects of quantum improvement (in the context of renormalization group improvement) on the affine theories of gravity.
Affine connection as a gravitational gauge field would encounter considerable changes after quantum improvement. The antiscreening feature of the running gravitational coupling manifests itself as some sort of anti gravity.

Although in general relativity the conditions of metricity and torsion--freedom are satisfied, quantum improvement can changes the scene.
We saw that simply improving the connection leads to the compactification of light cones. The trace of this compactification can be searched more in the context of the geodesic congruence behavior\cite{our4} and specially in the traversability of wormholes\cite{our5}.

On the other hand, the main feature of the connection improvement should be sought in the pure affine gravity theory, where the connection is considered as a dynamical object. To do so, we used the Schr\"{o}dinger--Eddington action and improved it.
The result is the classical Schr\"{o}dinger--Eddington equations modified by some quantum source terms.

The spherical symmetric vacuum solution of these equations is derived using perturbative methods. At this end, it is fruitful to have a comparison between the results of various methods of the quantum improvement for such a solution. 
In \cite{our2}, we have shown that the action improvement of metric theory leads to 
\begin{equation}
 g_{rr} \simeq -\frac{u}{u-1} - \frac{1}{\ln(u)}\ \ \ \ \textit{for small } u\ .
\end{equation}

\begin{figure}
    \centering
		\includegraphics[width=0.7\textwidth]{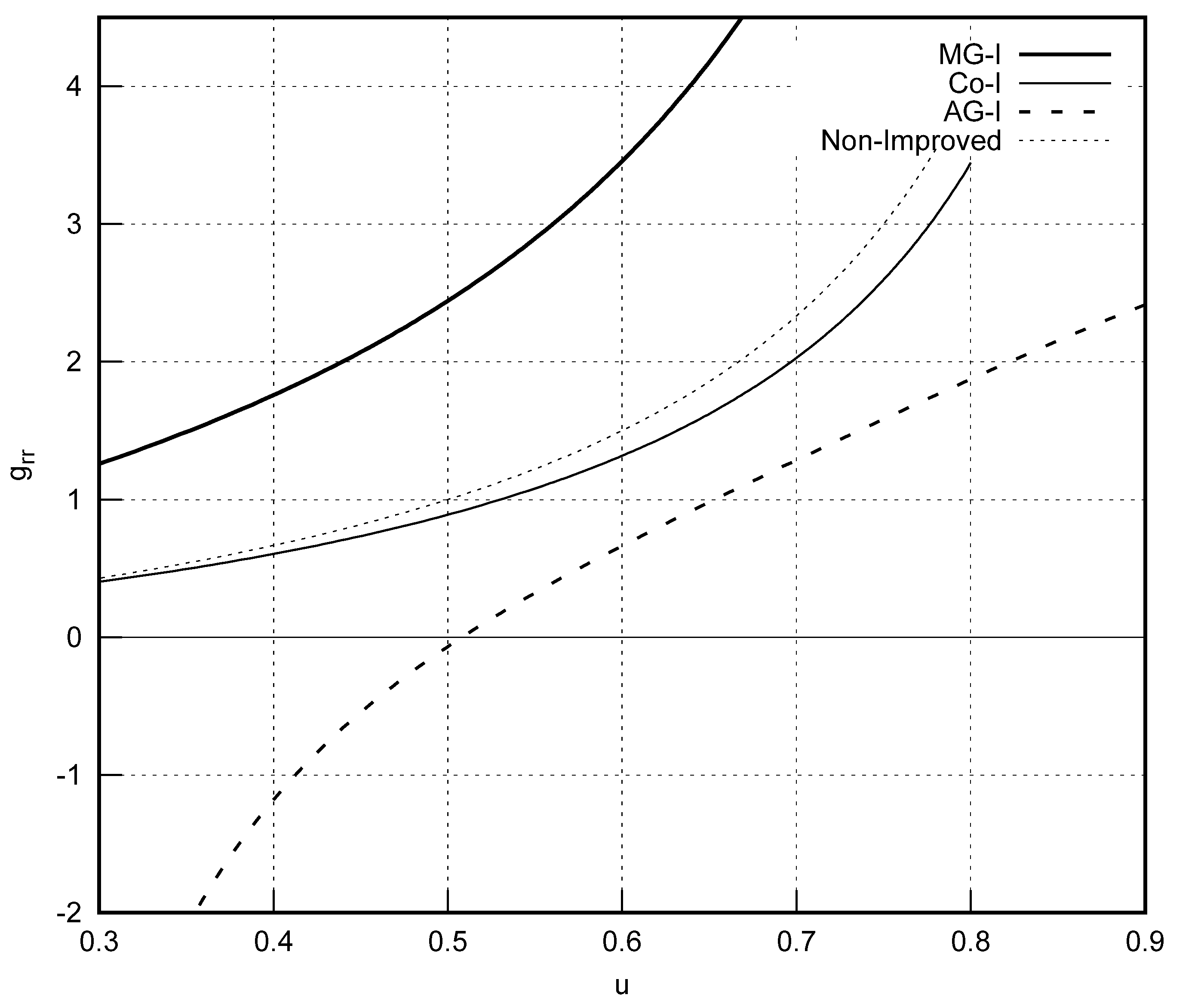}
	\caption{Comparison of the radial components of the improved metric for $\zeta = 0.1$. The improved metric gravity (MG-I), the improved connection (Co-I), the improved pure affine gravity (AG-I) and the classical Schwarzschild  are represented by the thick line, the thin line, the thick dashed line and the thin dashed line, respectively.}
	\label{grr-0.1}
\end{figure}

In figure \ref{grr-0.1}, we have compared the radial component of the metric for quantum improved metric gravity and affine theory (discussed here).
For the sake of comparison, the classical metric and the metric of connection improved in metric theory (section 2, equation \eqref{Im-rr}) are also presented.
As it is stated in the introduction, the most physical way of improvement seems to be affine improvement, because the affine is what carries the gravitational force. We see that affine improvement leads to significant changes near singularity.

\vglue1cm
\textbf{Acknowledgment:} This work is supported by a grant from Iran National Science Foundation (INSF).


\begin{thebibliography}{}
\bibitem{Schrodinger-B}{} E. Schr\"{o}dinger, \textit{Space--time Structure}, Cambridge University Press,
Cambridge (1950).
\bibitem{Eddington-B}{} A.S. Eddington, \textit{The Mathematical Theory of Relativity}, Cambridge University Press,
Cambridge (1924).
\bibitem{Kijowski}{} J. Kijowski, \href{https://doi.org/10.1007/BF00759646}{Gen. Rev. Grav. \textbf{9}, 857} (1978).
\bibitem{E-S}{} A. Einstein and E.G. Straus, \href{https://doi.org/10.2307/1969231}{Ann. Math. \textbf{47}, 731} (1946).
\bibitem{EM-GR1}{} N.J. Poplawski,  \href{https://doi.org/10.1142/S0217732307025662}{Mod. Phys. Lett. A \textbf{22},  2701} (2007).
\bibitem{EM-GR2}{} N.J. Poplawski,  \href{http://arxiv.org/abs/arXiv:0705.0351}{arXiv:0705.0351} (2007).
\bibitem{EM-GR3}{} N.J. Poplawski,  \href{https://doi.org/10.1142/S0217751X08039578}{IJMP A \textbf{23},  567} (2008).
\bibitem{EM-GR4}{} N.J. Poplawski,  \href{https://doi.org/10.1142/S0218271809014777}{IJMP D \textbf{18},  809} (2009).
\bibitem{PM}{} S. Weinberg, \textit{Gravitation and Cosmology: Principles and Applications of the General Theory of Relativity}, John Wiley \& Sons, New York (1972).
\bibitem{PM2}{} R.M. Wald, \textit{General Relativity}, The  University of Chicago Press, Chicago (1984).
\bibitem{MA1}{} F.W. Hehl, J.Dermott McCrea, E. W. Mielke and Y. Ne'eman, \href{https://doi.org/10.1016/0370-1573(94)00111-F}{Phys. Rept. \textbf{258}, 1-171} (1995).
\bibitem{MA2}{} M. Ferraris, M. Francaviglia and I. Volovich, \href{https://doi.org/10.1088/0264-9381/11/6/015}{Class. Quant. Grav. \textbf{11},  1505} (1994).
\bibitem{MA3}{} V. Tapia and M. Ujevic, \href{https://doi.org/10.1088/0264-9381/15/11/028}{Class. Quant. Grav. \textbf{15}, 3719} (1998).
\bibitem{MA4}{} T.P. Sotiriou and S. Liberati, \href{https://doi.org/10.1016/j.aop.2006.06.002}{Annals Phys. \textbf{322}, 935} (2007).
\bibitem{PA1}{} J.L. Cervantes--Cota and D.E. Liebscher,  \href{https://doi.org/10.1007/s10714-016-2103-9}{Gen. Rev. Grav. \textbf{48}, 108} (2016).
\bibitem{PA2}{} K. G\"{u}ltekin, \href{https://doi.org/10.1140/epjc/s10052-016-4015-y}{Eur. Phys. J. C \textbf{76}, 164} (2016).
\bibitem{Weinberg1}{}  S. Weinberg, in \textit{Understanding the Fundamental Constituents of Matter}, edited by A. Zichichi, Plenum Press, New York (1978).
\bibitem{Weinberg2}{} S. Weinberg, in \textit{General Relativity}, edited by S.W. Hawking and W. Isreal, Cambridge University Press, Cambridge (1979).
\bibitem{Becker}{}   D. Becker and M. Reuter, \href{https://doi.org/10.1007/JHEP07(2012)172}{JHEP \textbf{2012}, 172} (2012).
\bibitem{Niedermaier}{} M. Niedermaier and M. Reuter, \href{https://doi.org/10.12942/lrr-2006-5}{Liv. Rev. Rel. \textbf{9}, 5} (2006).
\bibitem{NGF}{} S. Weinberg, \href{https://doi.org/10.1103/PhysRevD.81.083535}{Phys. Rev. D \textbf{81}, 083535} (2010).
\bibitem{Reuter-1st}{}  M. Reuter, \href{https://doi.org/10.1103/PhysRevD.57.971}{Phys. Rev. D \textbf{57}, 971} (1998).
\bibitem{Reuter & Weyer}{} M. Reuter and  H. Weyer, \href{https://doi.org/10.1103/PhysRevD.69.104022}{Phys. Rev. D \textbf{69}, 104022} (2004).
\bibitem{Reuter & Saueressig} M. Reuter and F. Saueressig, \textit{Quantum gravity and the functional renormalization group: The road towards asymptotic safety}, Cambridge University Press, Cambridge (2019).
\bibitem{our3}{} R. Moti and A. Shojai, \href{https://doi.org/10.1016/j.physletb.2019.04.062}{Phys. Lett. B \textbf{793}, 313} (2019).
\bibitem{our4}{} R. Moti and A. Shojai, \href{https://doi.org/10.1103/PhysRevD.101.064013}{Phys. Rev. D \textbf{101}, 064013} (2020).
\bibitem{cos1}{} A. Bonanno and M. Reuter, \href{https://doi.org/10.1103/PhysRevD.65.043508}{Phys. Rev. D \textbf{65},  043508} (2002).
\bibitem{cos2}{} A. Platania, \href{https://doi.org/10.3389/fphy.2020.00188}{Front. Phys. \textbf{8}, 188} (2020).
\bibitem{cos3}{} A. Bonanno, A. Contillo and R. Percacci, \href{https://doi.org/10.1088/0264-9381/28/14/145026}{Class. Quant. Grav. \textbf{28}  145026} (2011).
\bibitem{our1}{} R. Moti and A. Shojai,  \href{https://doi.org/10.1140/epjc/s10052-017-5510-5}{Euro. Phys. J. C \textbf{78}, 32} (2018).
\bibitem{Bonanno & Reuter}{} A. Bonanno and M. Reuter, \href{https://doi.org/10.1103/PhysRevD.62.043008}{Phys. Rev. D \textbf{62}, 043008} (2000).
\bibitem{BH1}{} M. Reuter and E. Tuiran, \href{https://doi.org/10.1103/PhysRevD.83.044041}{Phys. Rev. D \textbf{83}, 044041} (2011).
\bibitem{BH2}{} K. Falls, F. Litim and A. Raghuraman, \href{https://doi.org/10.1142/S0217751X12500194}{Int. J. Mod. Phys. A \textbf{27}, 1250019} (2012).
\bibitem{BH3}{} A. Held, R. Gold and A. Eichhorn, \href{https://doi.org/10.1088/1475-7516/2019/06/029}{JCAP \textbf{1906}, 029 } (2019).
\bibitem{our2}{}  R. Moti and A. Shojai, \href{https://doi.org/10.1142/S0217751X20500165}{Int. J. Mod. Phys. A \textbf{35},  2050016}  (2020).
\bibitem{our5}{} R. Moti and A. Shojai, \href{https://doi.org/10.1103/PhysRevD.101.064013}{Phys. Rev. D \textbf{101}, 124042} (2020).
\bibitem{vac1}{} G. Gubitosi, C. Ripken and F. Saueressig, \href{https://doi.org/10.1007/s10701-019-00263-1}{Found. Phys. \textbf{49}, 972} (2019).
\bibitem{vac2}{} C. Pagani and M. Reuter, \href{https://doi.org/10.1016/j.aop.2019.167972}{Annals Phys. \textbf{411},  167972} (2019).
\bibitem{vac3}{} C. Pagani and M. Reuter, \href{https://doi.org/10.3389/fphy.2020.00214}{Front. Phys. \textbf{8}, 214} (2020).
\bibitem{Souma}{} W. Souma, \href{https://doi.org/10.1143/PTP.102.181}{Prog. Theor. Phys. \textbf{102}, 181} (1999).
\bibitem{Ad-1} F.W. Hehl, P. Heide, G.D. Kerlick and J.M. Nester, \href{https://doi.org/10.1103/RevModPhys.48.393}{Rev. Mod. Phy. \textbf{48}, 393} (1976).
\bibitem{Ad-2} I.L. Shapiro, \href{https://doi.org/10.1016/S0370-1573(01)00030-8}{Phys. Rept. \textbf{357},  113} (2002).
\bibitem{Ad-3} R.T. Hammond, \href{https://doi.org/10.1088/0034-4885/65/5/201}{Rept. Prog. Phys. \textbf{65},  599} (2002).
\bibitem{Ellis}{} G.F.R Ellis and J.P. Uzan, \href{https://doi.org/10.1119/1.1819929}{Am. J. Phys. \textbf{73}, 240} (2005).
\bibitem{Izadi}{} A. Izadi and A. Shojai, \href{https://doi.org/10.1088/0264-9381/26/19/195006}{Class. Quant. Grav. \textbf{29}, 195006} (2009).
\bibitem{Poisson}{} E. Poisson, \textit{A Relativist's Toolkit}, Cambridge University Press, Cambridge (2004).
\bibitem{Metric}{} J. Kijowski and R. Werpachowski, \href{https://doi.org/10.1016/S0034-4877(07)80001-2}{Rept. Math. Phys. \textbf{59}, 1} (2004).
\bibitem{Papapetrou}{} A. Papapetrou, \href{https://www.jstor.org/stable/20488492}{Proc. R. Ir. Acad. A \textbf{52}, 69} (1981).
\bibitem{Pawlowski}{} J. M. Pawlowski and D. Stock, \href{https://doi.org/10.1103/PhysRevD.98.106008}{Phys. Rev. D \textbf{98}, 106008} (2018).
\end{thebibliography}
\end{document}